\documentstyle[prc,aps]{revtex}

\begin{document}
\draft
\title{Faddeev description of two-hole one-particle motion
and the single-particle spectral function}  
\author{C.~Barbieri and W.~H.~Dickhoff}
\address{Department of Physics, Washington University,
         St.Louis, Missouri 63130}
\date{\today}
\maketitle

%%%%%%%%%%%%%%%%%%%%%%%%%%%%%%%%%%%%%%%%%%
\begin{abstract}
 The Faddeev technique is employed to address the problem of
describing the influence of both particle-particle and
particle-hole phonons on the single-particle self-energy.
 The scope of the few-body Faddeev equations is extended to
describe the motion of two-hole one-particle (two-particle one-hole)
excitations. This formalism allows to sum  both particle-particle and
particle-hole phonons, obtained separately in the Random Phase Approximation.
The appearance of spurious solutions for the present application of the
Faddeev method is related to the inclusion
of a consistent set of diagrams.
The formalism presented here appears practical for finite nuclei and
achieves a simultaneous inclusion of particle-particle
and particle-hole phonons to all orders while the spurious solutions
are properly eliminated. 
\end{abstract}
\pacs{PACS numbers:  21.10.Jx, {\bf 21.60.-n}, 21.60.Jz}
%%%%%%%%%%%%%%%%%%%%%%%%%%%%%

\section{Introduction}
\label{sec:introduction}

In recent years, the study of $(e,e'p)$ reactions has been one of the
most useful tools to probe correlations in nuclei.
Absolute spectroscopic factors have become available for many
closed-shell nuclei~\cite{diep,sick,Lapik} demonstrating that the
removal probability for nucleons from these systems is reduced
by about 35\% in comparison with the simple shell model.
The theoretical description of this reduction requires the inclusion of both
short-range and long-range correlations.
For nuclear matter a strength removal of about 15\% is obtained by
including short-range correlations~\cite{bv91}. 
For ${}^{16}{\rm O}$ the inclusion of short-range correlations leads to 
removal of single-particle (sp) strength of the order of 10\%~\cite{md94,rad94}.
The inclusion of long-range correlations for heavier nuclei like 
${}^{48}{\rm Ca}$ yields a qualitative description of the sp
strength distribution by including in the nucleon self-energy
the coupling to either low-lying collective
particle-hole (ph) or particle-particle (pp) phonons calculated in 
Random Phase approximation (RPA)~\cite{Rijsdijk}.
The additional depletion of about 10\% due to short-range correlations
for this nucleus leads to a reasonable quantitative agreement for the
largest fragments of the experimental strength distribution.

Theoretical calculations of hole spectroscopic factors for
$^{16}{\rm O}$ are not so successful.
The experimental spectroscopic strength~\cite{leus}
for the knockout of a proton from both the $p_{1/2}$ and
$p_{3/2}$ shells corresponds to about $60\%$.
The 10\% reduction due to short-range correlations is mostly compensated by
the proper inclusion of the center-of-mass motion which enhances the 
probability for $p$ removal by about 7\%~\cite{O16jast}.
Calculations based on the Green's function approach, including both long- and 
short-range correlations, yield about a $25\%$ reduction~\cite{GeurtsO16}.
These results still need to be corrected for the center-of-mass effect.
It is therefore fair to conclude that the present theoretical results
for ${}^{16}{\rm O}$ are still about 20\% away from the experimental data.
The importance of low-energy correlations 
is clearly  demonstrated by the results
of Ref.~\cite{GeurtsO16} and their proper inclusion is
therefore crucial for a complete understanding of this puzzle.
In the latter work the self-energy was obtained
including the effects of interactions between both pp and ph
excitations in the Tamm-Dancoff approximation (TDA).
In order to account for the coupling to collective excitations that are actually
observed in ${}^{16}{\rm O}$ it is necessary to at least consider 
an RPA description of the isoscalar negative parity states~\cite{czer}.
To account for the low-lying isoscalar positive parity states an even more
complicated treatment will be required.
Sizable collective effects are also present in the pp and hole-hole (hh)
excitations involving tensor correlations for isoscalar and pair correlations
for isovector states. 
Another argument to improve the description of the coupling
of sp states to low-lying collective excitations is provided by the lack
of fragmentation at low energy obtained in present theoretical 
studies~\cite{GeurtsO16} in disagreement with experimental data.

One of the goals of the present work is to account for the collectivity in the
ph and pp (hh) channels in a consistent way while including these excitations
at least at the RPA level. 
Since the observed fragmentation and depletion
of the sp strength in ${}^{16}{\rm O}$ is quite substantial, it is reasonable to
assume that these features are also important in the description of the
excitations that contribute to the self-energy.
This results in a self-consistent formulation
where the dressing of the nucleons is incorporated in the description
of the collective excitations that ultimately leads to the dressing itself.
For this reason the present work will be formulated using self-consistent
Green's functions (SCGF).
This type of self-consistency must also be considered in describing
pairing correlations in semi-magic nuclei~\cite{jyuan,wim99}.

A formalism in which both pp and ph phonons
are treated at the RPA level in the self-energy
was proposed in~\cite{Schuck1}. 
This work focused on the two-time two-particle one-hole (2p1h)
propagator and generated a formulation that reduces
to either including the pp-RPA or the ph-RPA phonons in the self-energy
when ph or pp vertices are omitted, respectively.
This expansion, however, was obtained using some drastic assumptions and
disregarding some of the constraints that arise when propagators
in different diagrams of the expansion are connected.
Related to this issue is the appearance of unphysical solutions for
the 2p1h propagator that have been discussed in~\cite{rijs96}.
Results in~\cite{rijs96} have therefore been obtained mostly for
the TDA treatment of the 2p1h propagator.
Moreover, this approximation was obtained by employing 
mean-field (single-pole) sp propagators.
In the present self-consistent treatment, which sums fully dressed
propagators, this approach is no longer possible.

To proceed with the inclusion of
both pp and ph collectivity in the nucleon self-energy
it is important to note that
the naive summation of diagrams containing both pp and ph
phonons leads to serious inconsistencies.
This approximation is
depicted in Fig.~\ref{fig:pp_ph_sf}. 
The last of the three diagrams on the
right hand side is already contained in each of the other two and must
therefore be subtracted to avoid double counting.  
This subtraction introduces spurious poles in the Lehmann representation of the
self-energy and generates meaningless solutions of the Dyson equation.
The minus sign in front this term may also prevent in some cases
the proper normalization of the spectroscopic amplitudes.
This feature can be understood by considering a possible solution
near such a spurious pole. 
The normalization is determined by the derivative of the
self-energy at this energy~\cite{wim99}
and will not yield a correct result on account of the
additional minus sign when the third diagram dominates.
In addition, each of the first two terms in Fig.~\ref{fig:pp_ph_sf}
ignores the Pauli correlations
between the freely propagating line and the quasi-particles forming the phonons,
as noted in~\cite{Rijsdijk}. 
In the present work a formalism is pursued which sums the
contribution of the pp and ph phonons to the self-energy to all orders
avoiding the subtraction of the second order diagram.
The treatment of Pauli correlations is improved over methods that employ
ph RPA phonons in the self-energy since all exchange terms at the 2p1h
level are consistently included.

 Other approaches have been proposed in the literature that attempt to extend
the nature of the phonon correlations included by performing massive summations
of diagrams~\cite{Dan90,Dan94,parquet}.
 Nevertheless, a consistent resummation of both pp and ph phonons to all
orders has not been achieved in these papers.
The main problem in pursuing such an infinite summation of diagrams for
the 2p1h propagator, which includes 
both pp and ph RPA correlations,  is related to
the fact that a two-body interaction can invert the
sense of propagation of only two lines (i.e. change at most two holes in two
particles and vice versa) while the third line continues to propagate
in the original direction.
In this way, a propagator depending on more than two times is generated.
It is therefore necessary first to consider an exact formulation
involving the four-time Green's function for the 2p1h propagator.
Direct application of four-time propagator equations
presumably will remain impractical for the forseeable future.
Appropriate approximations to this equation are therefore necessary
to construct the relevant two-times Green's
functions which contains the sought after correlations.
The scheme studied in this paper consists in computing the RPA
phonons in the pp and ph channels, separately, and then summing them to all
orders employing a Faddeev technique~\cite{Fadd1,glock}.

A nontrivial problem in the implementation of the Faddeev
equations is the appearance of spurious solutions~\cite{Glokle79,Evans}
which also have to be considered for the 2p1h propagator.
As pointed out in~\cite{Glokle99}, the spurious eigenstates are easy
to recognize for the few-body problem since they also diagonalize the
unperturbed hamiltonian. Their main features are that their eigenvalues
are known and that their wave-function amplitudes sum up to zero.
The situation is more complicated in the many-body problem
when the Faddeev technique is employed.
In particular, the fulfillment of closure relations for pp and ph
amplitudes is related to the behavior of the spurious Faddeev eigenstates.
Without a consistent treatment of this relation the spurious solutions
will mix with the physically meaningful ones.
Applying the Faddeev technique to the many-body problem, it
is important to solve for all physical solutions that contribute to
the self-energy.
Thus, it is necessary to develop a formalism in which the spurious solutions are 
correctly separated from the physical ones.
 
The practical implementation of the present Faddeev scheme is beyond the scope
of this paper.
The resulting set of equations require a great deal of
computational effort, especially when dressed propagators are employed.
Nevertheless, it appears that they can be solved in practice and results
using this formalism will be presented elsewhere.

In Sec.~\ref{sec:Rand4timesFd} we briefly describe the 
SCGF approach based on the Dyson equation and present the
exact Faddeev formalism for the four-times 2p1h propagator.
The construction of a consistent formulation for the two-time
2p1h propagator including the propagation of the pp and ph 
RPA phonons to all orders is presented in Sec.~\ref{sec:Faddeev}.
Although sofar only the 2p1h propagator has been mentioned, it should be
understood that the corresponding two-hole one-particle (2h1p) propagator
must be included in the calculation of the nucleon self-energy.
In the present work no coupling terms are considered 
that transform the 2p1h into the 2h1p propagators (or vice versa).
For this reason the same technique can be used for both propagators
and we will use the generic 2p1h to represent both.
In Sec.~\ref{sec:spurious} the appearance and treatment of spurious solutions
is discussed.
Some technical details are relegated to the appendices.
Conclusions are drawn in Sec.~\ref{sec:conclusions}.

%%%%%%%%%%%%%%%%%%%%%%%%%%%%%%%%%

\section{Self-Consistent Green's Function approach and 2p1h propagator}
\label{sec:Rand4timesFd}

\subsection{Self-energy and 2p1h propagator}

We consider a finite system of $A$ fermions interacting by means of
a two-body interaction $\hat{V}$.
As usual one may introduce an appropriate mean-field
potential $\hat{U}$ to localize the nucleons
and split the hamiltonian into an unperturbed one-body
part $\hat{H}_0 = \hat{T} + \hat{U}$ and a residual interaction
$\hat{H}_1 = \hat{V} - \hat{U}$.
Since we are mainly interested in low-lying bound states of
finite systems, we consider sp states with discrete quantum numbers.
As a basis, we choose the set of sp states $\{\alpha\}$ that
diagonalize $\hat{H}_0$ with corresponding eigenvalues
$\varepsilon^0_\alpha$.
The total hamiltonian can then be written as
 \begin{equation}
\hat{H}~=~\hat{H}_0 ~+~ \hat{H}_1
 ~=~\sum_{\alpha} \varepsilon^0_\alpha ~c^{\dag}_\alpha c_\alpha ~+~ 
   \left( {\textstyle {1\over 4}} \sum_{\alpha \beta \gamma \delta} 
              V_{\alpha \beta , \gamma \delta} 
                  ~c^{\dag}_\alpha ~c^{\dag}_\beta ~c_\delta ~c_\gamma 
                              ~-~ \sum_{\alpha \beta} U_{\alpha,\beta}
 ~c^{\dag}_\alpha c_\beta \right) \; ,
\label{eq:Hamiltonian}
\end{equation}
where $c^{\dag}_\alpha$ ($c_\alpha$) are the creation (destruction) operators
of a particle in the state $\alpha$, $V_{\alpha \beta , \gamma \delta}$ are
the antisymmetrized matrix elements of $V$, and $U_{\alpha,\beta}$
correspond to the matrix elements of $U$.

 The one-body propagator of the $A$-body system with ground state
$| \Psi^A_0 \rangle$ is defined as~\cite{fetwa,AAA}
\begin{equation}
g_{\alpha \beta}(\tau) ~=~ - i~\langle \Psi^A_0 | ~T [c_\alpha(\tau)  
                      c^{\dag}_\beta(0) ]~| \Psi^A_0 \rangle \; , 
\label{eq:fullg_time}
\end{equation}
where $c^{\dag}_\alpha(t)$ and $c_\alpha(t)$ now correspond
to operators in the Heisenberg picture.
In the Lehmann representation, all the eigenvalues of the excited states
of the systems with $A+1$ and $A-1$ particles appear, as well as their
spectroscopic amplitudes for transitions to those states
that are relevant for comparison with experimental data.

The propagator $g_{\alpha \beta}(\tau)$ can be obtained as the sum of
an infinite set of diagrams, built from interaction vertices and unperturbed
sp propagators $g_{\alpha\beta}^{(0)}$ corresponding to $\hat{H}_0$.
 In the nuclear case, a strong coupling exists between the sp degree of
freedom and both collective low-lying states as well as high-lying states.
The latter coupling is related to the strong short-range repulsion
in the nuclear force.
The resulting fragmentation of the sp strength, as observed in experimental
data, suggests that this feature must already be included in the description
of these couplings.
For this reason self-consistent one-body propagators need to be considered
in the construction of the nucleon self-energy.
This self-consistency feature also emerges in an exact formulation, involving 
the coupling to two-, three- and $A$-body propagators, which can be
derived using the equation of motion method~\cite{masch}.
In short, this means that for the nuclear case one needs
to develop the perturbation theory in
terms of the dressed propagator~(\ref{eq:fullg_time})
approximated in an appropriate way.

The approach we use here consists in computing $g_{\alpha \beta}(\tau)$
as a solution of the Dyson equation
\begin{equation}
g_{\alpha \beta}(\tau) ~=~ g^{(0)}_{\alpha \beta}(\tau) ~+~ 
                              g^{(0)}_{\alpha \gamma}(\tau - t_1)
                              \Sigma^{\star}_{\gamma \delta}(t_1 - t_2)
                              g_{\delta \beta}(t_2)  \; ,
\label{eq:Dyson}
\end{equation}
where $\Sigma^{\star}_{\alpha \beta}(\tau)$ is the irreducible self-energy.
Here and in the following, we employ the convention of summing over
all repeated indices and integrate from $-\infty$
to +$\infty$ over all repeated time variables, unless specified otherwise.
 
By considering the equation of motion for $g_{\alpha \beta}(\tau)$, one
obtains that
$\Sigma^{\star}_{\alpha \beta}(\tau)$ can be written as the sum of two terms
\begin{equation}
 \Sigma^{\star}_{\alpha \beta}(\tau) ~=~  \Sigma^{HF}_{\alpha \beta} 
   ~+~ V_{\alpha \lambda , \mu \nu} ~
     R_{\mu \nu \lambda , \gamma \delta \varepsilon}(\tau^- , \tau , \tau^+ ;
 0^+ , 0 , 0^- )
         ~ V_{\gamma \delta , \beta \varepsilon}  \; ,
\label{eq:Sigma1}
\end{equation}
where $\Sigma^{HF}_{\alpha \beta}$ represents the (time independent)
Hartree-Fock part of the self-energy, which can be computed from the
solution $g_{\alpha \beta}(\tau)$ itself.
The 2p1h propagator $R$, appearing in the
last term of Eq.~(\ref{eq:Sigma1}), contains the sum of all so-called
one-particle irreducible diagrams
which cannot be separated by cutting a single line.
These terms are included in the 2p1h Green's function, 
$g^{2p1h}_{\mu \nu \lambda , \alpha \beta \gamma}$, defined below. 
The relation between $R$ and $g^{2p1h}$ is given by~\cite{Win72}
 \begin{eqnarray}
  \lefteqn{
 R_{\mu \nu \lambda , \alpha \beta \gamma}(t_1 , t_2 , t_3 ; t_4 , t_5 , t_6) }
    \hspace{.5in} & &
\nonumber   \\
    & & = g^{2p1h}_{\mu \nu \lambda , \alpha \beta \gamma}
                     (t_1 , t_2 , t_3 ; t_4 , t_5 , t_6)
         ~-~ g^{II}_{\mu \nu ,\lambda \eta}(t_1 , t_2 ; t_3 , t') \;
              g^{-1}_{\eta \sigma}(t' - t'') \;
              g^{II}_{\gamma \sigma , \alpha \beta}(t_6 , t'' ; t_4 , t_5) \; , 
\label{eq:Rvsg}
\end{eqnarray}
in which $g^{-1}_{\eta \sigma}$ is the inverse of the one-body Green's
function~(\ref{eq:fullg_time}) and $g^{II}$ and $g^{2p1h}$ are
the 4- and 6- point Green's functions defined as
\begin{equation}
     g^{II}_{\alpha \beta ,\gamma \delta}(t_1 , t_2 ; t_3 , t_4) ~=~
          - i ~ \langle \Psi^A_0 |
              ~T [c_\beta(t_2) c_\alpha(t_1) c^{\dag}_\gamma(t_3) c^{\dag}_\delta(t_4) ]
              ~| \Psi^A_0 \rangle \;  
\label{eq:g2}
\end{equation}
and
\begin{equation}
   g^{2p1h}_{\mu \nu \lambda , \alpha \beta \gamma}(t_1 , t_2 , t_3 ; t_4 , t_5 , t_6) ~=~
          - i ~ \langle \Psi^A_0 |
              ~T [c^{\dag}_\lambda(t_3) c_\nu(t_2) c_\mu(t_1) 
              c^{\dag}_\alpha(t_4) c^{\dag}_\beta(t_5) c_\gamma(t_6) ]
              ~| \Psi^A_0 \rangle \;  ,
\label{eq:g3}
\end{equation}
respectively.

 The propagator $R_{\mu \nu \lambda , \alpha \beta \gamma}$
is the solution of the following equation which has a similar form
as the Bethe-Salpeter equation for pp and ph propagators
\begin{eqnarray}
\lefteqn{
 R_{\mu \nu \lambda , \alpha \beta \gamma}(t_1 , t_2 , t_3 ; t_4 , t_5 , t_6)}
    \hspace{.5in} & &
\nonumber \\
  &=&  g_{\mu \alpha}(t_1 - t_4) g_{\nu \beta}(t_2 - t_5) g_{\gamma \lambda}(t_6 - t_3)
     ~-~ g_{\nu \alpha}(t_2 - t_4) g_{\mu \beta}(t_1 - t_5) g_{\gamma \lambda}(t_6 - t_3)
\nonumber \\
  & & + g_{\mu \mu'}(t_1 - t_1') g_{\nu \nu'}(t_2 - t_2') g_{\lambda' \lambda}(t_3' - t_3)
\nonumber \\  
  & & ~ \times \; K_{\mu' \nu' \lambda' , \alpha' \beta' \gamma'}(t_1' , t_2' , t_3' ; t_4' , t_5' , t_6')
    \;  R_{\alpha' \beta' \gamma' , \alpha \beta \gamma}(t_4' , t_5' , t_6' ; t_4 , t_5 , t_6) \; ,
\label{eq:Rbt}
\end{eqnarray}
which is shown in Fig.~\ref{fig:BS2p1h} in terms of Feynman diagrams.
The interaction vertex, also shown in Fig.~\ref{fig:BS2p1h}, is given by
\begin{eqnarray}
\lefteqn{
 K_{\mu \nu \lambda , \alpha \beta \gamma}(t_1 , t_2 , t_3 ; t_4 , t_5 , t_6) }
    \hspace{.5in} & &
\nonumber \\
  &=& K^{(ph)}_{\nu \lambda , \beta \gamma}(t_2 , t_3 ; t_5 , t_6) g^{-1}_{\mu \alpha}(t_1 - t_4) ~+~
  K^{(ph)}_{\mu \lambda , \alpha \gamma}(t_1 , t_3 ; t_4 , t_6) g^{-1}_{\nu \beta}(t_2 - t_5)
\nonumber \\
 & & ~+~ K^{(pp)}_{\mu \nu , \alpha \beta}(t_1 , t_2 ; t_4 , t_5) g^{-1}_{\gamma \lambda}(t_6 - t_3)
 ~+~K^{(pph)}_{\mu \nu \lambda , \alpha \beta \gamma}(t_1 , t_2 , t_3 ; t_4 , t_5 , t_6)  \; .
\label{eq:Rbtvertex}
\end{eqnarray}
In Eq.~(\ref{eq:Rbtvertex}), $K^{(pp)}$ and $K^{(ph)}$ represent the pp and
ph irreducible vertices while $K^{(pph)}$ is the 2p1h irreducible vertex.
It should be noted that in Eq.~(\ref{eq:Sigma1}) the propagator
$R_{\mu \nu \lambda , \alpha \beta \gamma}$ is only required at
two times and therefore its complete knowledge, as given by~(\ref{eq:Rvsg}), is
not necessary to solve the Dyson equation.
On the other hand, the dependence on the time variables $t_1$, $t_2$ and $t_3$
is employed in the Bethe-Salpeter Eq.~(\ref{eq:Rbt}), thus requiring that at
least a 4-times object be employed to solve for the 2p1h motion
exactly.

 Equations~(\ref{eq:Dyson}),~(\ref{eq:Sigma1}) and~(\ref{eq:Rbt}) together
form a set of coupled equations, where the same propagator,
which solves the Dyson equation~(\ref{eq:Dyson}), appears as input in the
Bethe-Salpeter equation~(\ref{eq:Rbt}).
If the irreducible vertices $K^{(pp)}$, $K^{(ph)}$ and $K^{(pph)}$ are also
expressed in terms of the $g_{\alpha \beta}(\tau)$, then
Eqs.~(\ref{eq:Dyson}) and~(\ref{eq:Rbt}) will generate a self-consistent
expansion.
Obviously, Eq.~(\ref{eq:Rbt}) and the irreducible vertex~(\ref{eq:Rbtvertex})
represent the exact solution for $R$ and therefore require a suitable
approximation.

\subsection{Faddeev-Bethe-Salpeter equations}

 Eq.~(\ref{eq:Rbt}) can be reduced to a set of coupled equations in a way
similar to the method proposed by Faddeev to solve the three-body
problem~\cite{Fadd1,JoachBk}.
The inclusion of pp and ph RPA phonons in a consistent way requires
this Faddeev approach since it provides a natural framework for correctly
iterating quantities that have already been summed to all orders
like these RPA phonons.
In the present work we will neglect the contribution of
the irreducible $K^{(pph)}$ term in Eq.~(\ref{eq:Rbtvertex}) since
it leads to the coupling of higher order particle-hole terms than already
considered in the following.
We will therefore require only three Faddeev components.
Following standard notation in the literature~\cite{glock}, 
$R^{(i)}_{\mu \nu \lambda , \alpha \beta \gamma}$ will represent 
the component related to all diagrams ending with a vertex between legs $j$
and $k$ with $(i,j,k)$ cyclic permutations of $(1,2,3)$.
We will employ the convention in which the third leg propagates in the opposite
direction with respect to the first two.
The Faddeev components $R^{(i)}$ can be written in terms of the 2p1h
propagator $R$ and the contribution of the three dressed but noninteracting
sp propagators.
This definition is given in detail here for all three components, omitting
explicit reference to the time variables for convenience of notation
\begin{mathletters}
\label{eq:faddcmpall}
\begin{equation}
    R^{(1)}_{\mu \nu \lambda , \alpha \beta \gamma} ~=~
     g_{\nu \epsilon} g_{\rho \lambda} ~
        K^{(ph)}_{\epsilon \rho,\eta \sigma} ~
          R_{\mu \eta \sigma , \alpha \beta \gamma}
     ~+~ \frac{1}{2} \left(
                       g_{\mu \alpha} \; g_{\nu \beta} \; g_{\gamma \lambda}
                     - g_{\nu \alpha} \; g_{\mu \beta} \; g_{\gamma \lambda}
                     \right) \; ,
\label{eq:faddcmp1}
\end{equation}
\begin{equation}
    R^{(2)}_{\mu \nu \lambda , \alpha \beta \gamma} ~=~
     g_{\mu \epsilon} g_{\rho \lambda} ~
       K^{(ph)}_{\epsilon \rho,\eta\sigma} ~
         R_{\eta \nu \sigma , \alpha \beta \gamma}
     ~+~ \frac{1}{2} \left(
                       g_{\mu \alpha} \; g_{\nu \beta} \; g_{\gamma \lambda}
                     - g_{\nu \alpha} \; g_{\mu \beta} \; g_{\gamma \lambda}
                     \right) \; ,
\label{eq:faddcmp2}
\end{equation}
\begin{equation}
    R^{(3)}_{\mu \nu \lambda , \alpha \beta \gamma} ~=~
     g_{\mu \epsilon} g_{\nu \rho} ~
       K^{(pp)}_{\epsilon \rho,\eta\sigma} ~
         R_{\eta \sigma \lambda , \alpha \beta \gamma}
     ~+~ \frac{1}{2} \left(
                       g_{\mu \alpha} \; g_{\nu \beta} \; g_{\gamma \lambda}
                     - g_{\nu \alpha} \; g_{\mu \beta} \; g_{\gamma \lambda}
                     \right) \; .
\label{eq:faddcmp3}
\end{equation}
%
%\begin{equation}
%    R^{(3)}_{\mu \nu \lambda , \alpha \beta \gamma}(t_1 , t_2 , t_3 ; t_4, t_5 , t_6) ~=~
%     g_{\mu \epsilon}(t_1 - t_1') g_{\nu \rho}(t_2 - t_2') 
%        K^{pp}_{\epsilon \rho,\eta\sigma}(t_1' - t_4' ; t_2' - t_5')
%           R_{\eta \sigma \lambda , \alpha \beta \gamma}(t_4', t_5' , t_6' ; t_4, t_5 , t_6) 
%     ~+~ \frac{1}{2} \ g^{-1}_{\mu \alpha} g^{-1}_{\nu \beta} g^{-1}_{\gamma \lambda} \; ,
%\label{eq:faddcmp3}
%\end{equation}
\end{mathletters}
The factor $\frac{1}{2}$ in Eqs.~(\ref{eq:faddcmpall}) properly takes
into account the exchange symmetry between the parallel lines in the
Faddeev equations.
 With these definitions the full propagator~(\ref{eq:Rvsg}) is given by
\begin{equation}
    R_{\mu \nu \lambda , \alpha \beta \gamma} ~=~
   \sum_{i=1,2,3}     R^{(i)}_{\mu \nu \lambda , \alpha \beta \gamma}
     ~-~ \frac{1}{2} \left( g_{\mu \alpha} \; g_{\nu \beta} \; g_{\gamma \lambda}
        - g_{\nu \alpha} \; g_{\mu \beta}  \; g_{\gamma \lambda} \right) \; .
\label{eq:faddfullR}
\end{equation}
The Faddeev equations now take the following form
\begin{eqnarray}
    R^{(i)}_{\mu \nu \lambda , \alpha \beta \gamma} &=&
          \frac{1}{2} \left( g_{\mu \alpha} \; g_{\nu \beta} \; g_{\gamma \lambda}
               - g_{\nu \alpha} \; g_{\mu \beta}  \; g_{\gamma \lambda} \right)
\nonumber \\ 
   & & +~  g_{\mu \mu'} \; g_{\nu \nu'} \; g_{\lambda' \lambda} ~
            \Gamma^{(i)}_{\mu' \nu' \lambda' , \mu'' \nu'' \lambda''} ~
            (  R^{(j)}_{\mu'' \nu'' \lambda'' , \alpha \beta \gamma} ~+~
               R^{(k)}_{\mu'' \nu'' \lambda'' , \alpha \beta \gamma} ) \; ,
                ~ ~ i = 1, 2, 3
\label{eq:BSFaddeq}
\end{eqnarray}
where the $\Gamma^{(i)}_{\mu \nu \lambda , \alpha \beta \gamma}$ vertices
obey the following symmetry relations and are defined by 
\begin{mathletters}
\label{eq:G2vsBS}
\begin{equation}
   \Gamma^{(1)}_{\mu \nu \lambda , \alpha \beta \gamma}
                                       (t_1 , t_2 , t_3 ; t_4 , t_5 , t_6) =
   \Gamma^{(2)}_{\nu \mu \lambda , \beta \alpha \gamma}
                                       (t_2 , t_1 , t_3 ; t_5 , t_4 , t_6) =
  g^{-1}_{\mu \alpha}(t_1 - t_4) ~
 \tilde{\Gamma}^{(ph)}_{\nu \lambda , \beta \gamma}(t_2 , t_3 ; t_5 , t_6) \; ,
\label{eq:G2vsBS_ph}
\end{equation}
\begin{equation}
   \Gamma^{(3)}_{\mu \nu \lambda , \alpha \beta \gamma}
                                       (t_1 , t_2 , t_3 ; t_4 , t_5 , t_6) =
   \Gamma^{(3)}_{\nu \mu \lambda , \beta \alpha \gamma}
                                       (t_2 , t_1 , t_3 ; t_5 , t_4 , t_6) =
    g^{-1}_{\gamma \lambda}(t_6 - t_3) ~
     \tilde{\Gamma}^{(pp)}_{\mu \nu , \alpha \beta}(t_1 , t_2 ; t_4 , t_5) \; .
\label{eq:G2vsBS_pp}
\end{equation}
\end{mathletters}
The gamma matrices $\tilde{\Gamma}^{(pp)}$ and
$\tilde{\Gamma}^{(ph)}$ are the four-point functions that
solve the Bethe-Salpeter equation for the
pp and ph motion.
These vertex functions contain the pp and ph phonons and can be written as
\begin{mathletters}
\label{eq:BetheSalpeter}
\begin{eqnarray}
 \lefteqn{
  \tilde{\Gamma}^{(pp)}_{\gamma \delta , \alpha \beta}(t_1 , t_2 ; t_3 , t_4) ~=~
   K^{(pp)}_{\gamma \delta , \alpha \beta}(t_1 , t_2 ; t_3 , t_4) }
     \hspace{.5in} & &
\nonumber \\
  &+&
    \tilde{\Gamma}^{(pp)}_{\gamma \delta , \mu \nu}(t_1 , t_2 ; t_1' , t_2') ~
     g_{\mu \eta}(t_1' - t_3') \; g_{\nu \sigma}(t_2' - t_4') ~
      K^{(pp)}_{\eta \sigma , \alpha \beta}(t_3' , t_4' ; t_3 , t_4) \; ,
\label{eq:ppBS}
\end{eqnarray}
\begin{eqnarray}
 \lefteqn{
  \tilde{\Gamma}^{(ph)}_{\gamma \delta , \alpha \beta}(t_1 , t_2 ; t_3 , t_4) ~=~
   K^{(ph)}_{\gamma \delta , \alpha \beta}(t_1 , t_2 ; t_3 , t_4) }
     \hspace{.5in} & &
\nonumber \\ 
  &+&
    \tilde{\Gamma}^{(ph)}_{\gamma \delta , \mu \nu}(t_1 , t_2 ; t_1' , t_2') ~
     g_{\mu \eta}(t_1' - t_3') \; g_{\sigma \nu}(t_4' - t_2') ~
      K^{(ph)}_{\eta \sigma , \alpha \beta}(t_3' , t_4' ; t_3 , t_4) \; .
\label{eq:phBS}
\end{eqnarray}
\end{mathletters}

%%%%%%%%%%%%%%%%%%%%%%%%%%%%%%%%%

\section{Approximate Faddeev equations for 2p1h motion}
\label{sec:Faddeev}

Apart from neglecting the $K^{(2p1h)}$ vertex,
Eq.~(\ref{eq:BSFaddeq}) is otherwise a complete
equation for the 2p1h propagator.
This general equation involves quantities which depend on several times
and is therefore too complex to be solved numerically.
In order to construct a manageable approximation scheme that includes the
relevant physical ingredients two simplifications will be considered in this 
section.
The first one involves the restriction to two-time pp and ph
vertices that include the respective RPA contributions in these channels.
This approximation is the minimum step that maintains the simultaneous inclusion
of both pp and ph collective low-lying excitations in describing the sp
propagator. 
Second, it is necessary to simplify Eq.~(\ref{eq:BSFaddeq}) to include
only two-times Green's functions.
This procedure no longer allows the inversion of the
the propagation direction of all three lines together.
As a result, the Faddeev equations split up in two
separate expansions for the 2p1h and the 2h1p components.
Although the hole spectral function is of primary interest for comparison with 
experimental data, it must be
stressed that both 2p1h and 2h1p components are needed to generate the
self-consistent solution for the sp propagator.
Since the formalism involved is the same for both components, we will
describe only the forward-going (2p1h) expansion. The equations for
the 2h1p case are completely analogous.

\subsection{Faddeev equations} 

 To construct the present approximation scheme, it is more convenient
to use the energy representation.
The corresponding Lehmann representation of the
sp propagator~(\ref{eq:fullg_time}) is given by
\begin{equation}
 g_{\alpha \beta}(\omega) ~=~ 
 \sum_n  \frac{ \left( {\cal X}^{n}_{\alpha} \right)^* \;{\cal X}^{n}_{\beta} }
                       {\omega - \varepsilon^{+}_n + i \eta }  ~+~
 \sum_k \frac{ {\cal Y}^{k}_{\alpha} \; \left( {\cal Y}^{k}_{\beta} \right)^*  }
                       {\omega - \varepsilon^{-}_k - i \eta } \; ,
\label{eq:fullg}
\end{equation}
where ${\cal X}^{n}_{\alpha} = {\mbox{$\langle {\Psi^{A+1}_n} \vert $}}
 c^{\dag}_\alpha {\mbox{$\vert {\Psi^A_0} \rangle$}}$%
~(${\cal Y}^{k}_{\alpha} = {\mbox{$\langle {\Psi^{A-1}_k} \vert $}}
 c_\alpha {\mbox{$\vert {\Psi^A_0} \rangle$}}$) are the
spectroscopic amplitudes for the excited states of a system with
$A+1$~($A-1$) particles and the poles $\varepsilon^{+}_n = E^{A+1}_n - E^A_0$%
~($\varepsilon^{-}_k = E^A_0 - E^{A-1}_k$) correspond to the excitation energies
with respect to the $A$-body ground state.
In Eq.~(\ref{eq:fullg}) and in the following, we use the indices $n$
and $k$ to enumerate the fragments associated with the one-particle and 
one-hole excitations, respectively.

Employing the bare interaction
$V_{\alpha \beta , \gamma \delta}$ for the vertices $K^{(pp)}$ and $K^{(ph)}$,
the Bethe-Salpeter equations~(\ref{eq:BetheSalpeter}) reduce to the usual
dressed RPA (DRPA) equations~\cite{schuckRPA,DRPApaper}. The solutions
of these equations depend only on two times.
These pp and ph phonons correspond to the dressed version of the phonons that
are considered in Ref.~\cite{Rijsdijk} (see also Fig.~\ref{fig:pp_ph_sf}).
These excitations describe
the correlations that we aim to iterate to all orders 
and, subsequently, to include in the self-energy as explained
in the introduction.
These DRPA solutions can then be substituted in Eqs.~(\ref{eq:G2vsBS})
to generate the $\Gamma^{(i)}$ matrices to be used in the Faddeev expansion.
Both the forward-
and backward-going components of the DRPA solutions are included
into the expansion as illustrated in Fig.~\ref{fig:Gammai_vs_DRPAS}.
This is crucial in order to eliminate the spurious solutions of the
Faddeev equations as will be explained in Sec.~\ref{sec:spurious} and
Appendix~\ref{app:spurious}.

 The working expression for the $\Gamma^{(i)}$ matrices, which depends on only
two times (or equivalently one energy), is given in some detail in
Appendix~\ref{app:Gboxes}.
 Here we only need to stress that the resulting $\Gamma^{(i)}$'s 
cannot invert the freely propagating line from hole to particle or
vice versa, {\em i.e.} they cannot connect the 2p1h amplitudes with
the 2h1p ones.
For this reason, the pp and ph phonons will be summed only in
one time direction in a TDA way contributing separately to the 2p1h
and 2h1p propagators.
The reader may notice that two contributions of the type shown in
Fig.~\ref{fig:bfflip} can connect the 2p1h and the 2h1p propagators.
The inclusion of such terms leads to the simultaneous propagation of two
phonons which requires an extension of the approximation presented in this paper.
Since these terms are expected to contribute only in higher order,
we will neglect them in the following.
We note that the collective RPA correlations in the pp and ph channels have
already been computed through Eqs.~(\ref{eq:BetheSalpeter}) and therefore 
remain properly included in our approximation.

 The remaining complication, related to the use of dressed propagators,
concerns the interactions vertices~(\ref{eq:G2vsBS}).
As explained in Appendix~\ref{app:Gboxes}, the $\Gamma^{(i)}$ and the
propagators $R^{(i)}$ need to be redefined in such a way that their matrix
elements also depend on the indices ($n$, $n'$, $k$), which label the
fragments of the propagators.
This implies that the eigenvalue equations will involve
summations on both the sp indices ($\alpha$, $\beta$, $\gamma$) and the
ones corresponding to the fragmentation, ($n_\alpha$, $n_\beta$, $k_\gamma$).
 The 2p1h propagator and its Faddeev components, as defined
in Eqs.~(\ref{eq:Rvsg}) and~(\ref{eq:faddcmpall}), are recovered only at
the end by summing the solutions over all values
of ($n_\alpha$, $n_\beta$, $k_\gamma$) and~($n_\mu$,~$n_\nu$,~$k_\lambda$).

 Putting together all the above considerations, the resulting approximation to
the Faddeev equations~(\ref{eq:BSFaddeq}) can be rewritten in a 
way where all the propagators involved depend only on one energy variable
(or two time variables).
 The forward-going part of this expansion can be written as follows
\begin{eqnarray}
  \lefteqn{
  R^{(i)}_{\mu    n_\mu    \nu   n_\nu   \lambda k_\lambda , 
           \alpha n_\alpha \beta n_\beta \gamma  k_\gamma}(\omega) }
        \hspace{.5in} & &
\nonumber  \\
   &=&  \frac{1}{2}\left( 
      {G^0}^>_{\mu    n_\mu    \nu   n_\nu   \lambda k_\lambda , 
               \alpha n_\alpha \beta n_\beta \gamma  k_\gamma}(\omega)
    - {G^0}^>_{\nu   n_\nu   \mu    n_\mu    \lambda k_\lambda ,
               \alpha n_\alpha \beta n_\beta \gamma  k_\gamma}(\omega) \right)
\nonumber  \\
  & & +~ {G^0}^>_{\nu  n_\nu   \mu  n_\mu   \lambda  k_\lambda ,
                  \mu' n_\mu'  \nu' n_\nu'  \lambda' k_\lambda'}(\omega) ~
    \Gamma^{(i)}_{\nu'  n_\nu'  \mu'  n_\mu'   \lambda'  k_\lambda' ,
                  \mu'' n_\mu'' \nu'' n_\nu''  \lambda'' k_\lambda''}(\omega) ~
\nonumber  \\
  & & ~ ~ ~ ~ ~ ~ ~ ~ \times ~
  \left( R^{(j)}_{\mu''    n_\mu''    \nu''   n_\nu ''  \lambda'' k_\lambda'' , 
                  \alpha n_\alpha \beta n_\beta \gamma  k_\gamma}(\omega) ~+~
         R^{(k)}_{\mu''    n_\mu''    \nu''   n_\nu ''  \lambda'' k_\lambda'' , 
                  \alpha n_\alpha \beta n_\beta \gamma  k_\gamma}(\omega)
       \right)  \; , ~ ~ i=1,2,3 ~ .
\label{eq:FaddTDA}
\end{eqnarray}
 In Eq.~(\ref{eq:FaddTDA}), ${G^0}^>$ is the forward-going part of the 2p1h
propagator for three dressed but noninteracting lines.
 Using the notations introduced after Eq.~(\ref{eq:fullg}) we have
\begin{equation}
  {G^0}^>_{\mu    n_\mu    \nu   n_\nu   \lambda k_\lambda , 
           \alpha n_\alpha \beta n_\beta \gamma  k_\gamma}(\omega) ~=~
        \delta_{n_\mu , n_\alpha} \;
        \delta_{n_\nu , n_\beta} \;
        \delta_{k_\lambda , k_\gamma} ~
     \frac{ \left( 
         {\cal X}^{n_\mu}_{\mu}
         {\cal X}^{n_\nu}_{\nu}
         {\cal Y}^{k_\lambda}_{\lambda}
            \right)^* \; 
         {\cal X}^{n_\alpha}_{\alpha}
         {\cal X}^{n_\beta}_{\beta}
         {\cal Y}^{k_\gamma}_{\gamma} }
    { \omega - ( \varepsilon^+_{n_\alpha} + \varepsilon^+_{n_\beta} -
                                     \varepsilon^-_{k\gamma} ) + i \eta } \; .
\label{eq:G0fw}
\end{equation}

 Eqs.~(\ref{eq:FaddTDA}), together with the $\Gamma^{(i)}$'s given
in Appendix~\ref{app:Gboxes}, approximate the general 
``Faddeev-Bethe-Salpeter'' expansion to a tractable set of equations
involving only two-time objects.
It is important to note that these equations are still expressed in terms of the
self-consistent solution $g_{\alpha \beta}(\omega)$ and include both pp
and ph~RPA phonons in a correct way. 
 Thus they maintain all the features relevant for the physics we aim to
describe.

\subsection{Faddeev amplitudes}
\label{sub:xamp}

 Eqs.~(\ref{eq:FaddTDA}) involve the use of propagators
depending on a large number of indices. As a consequence, the dimension of the
problem could easily grow up to a point where no practical application is
feasible for a real system.
 This difficulty can be overcome by introducing a new set of spectroscopic
amplitudes which depend only on the indices labelling the particle and hole
fragments~($n$, $n'$, $k$)~\cite{Smain,RipkaBk}.
Thereby the problem is reexpressed by changing from the basis of sp
states $\{\alpha\}$, used in definitions~(\ref{eq:Rvsg})
and~(\ref{eq:faddcmpall}), to a new formulation constructed in terms
of the fragments labeled by $\{n,k\}$.
 This procedure also allows to rewrite the eigenvalue and normalization
conditions corresponding to Eqs.~(\ref{eq:FaddTDA}) in a more concise way.
 As long as the interaction elements $V_{\alpha \beta , \gamma \delta}$ are
energy independent, all the solutions can then be obtained through a
single diagonalization.
This approach is particularly satisfactory from a physical point of view since the
equations reflect the mixing of the 2p1h states
represented by the ($n$, $n'$, $k$) fragments.
This new formulation does not introduce any further
approximation.
Nevertheless, since it appears relevant for a practical
solution of the problem, we will describe it in the following.

 The Lehmann representation of the Faddeev components~
$R^{(i)}_{\mu n_\mu \nu  n_\nu \lambda k_\lambda ,
                     \alpha n_\alpha \beta n_\beta \gamma k_\gamma}$,
contains all the poles $\varepsilon^{Fd}_m$ of the 2p1h
propagator, each with its own residue.
One obtains for these components
\begin{equation}
 R^{(i)}_{\mu    n_\mu    \nu   n_\nu    \lambda k_\lambda ,
          \alpha n_\alpha \beta n_\beta \gamma  k_\gamma} ~=~ \sum_{m}  
      \frac{ \left(
                \beta^{(i),m}_{\mu n_\mu \nu  n_\nu \lambda k_\lambda}
             \right)^* 
           ~ b ^m_{\alpha n_\alpha \beta n_\beta \gamma k_\gamma} }
              {  \omega - \varepsilon^{Fd}_m + i \eta }
       ~+~  R^{(i)}_{free} ( \omega )  \; ,
 \label{eq:Fd_comp_Lehrp}
\end{equation}
 where the superscript $m$ labels the solutions of Faddeev
equations.
 In Eq.~(\ref{eq:Fd_comp_Lehrp}), $R^{(i)}_{free}$ represent
components containing
the same poles as ${G^0}^>$~(\ref{eq:G0fw}).
The sum of these terms cancels exactly the contribution of the three freely
propagating lines in Eq.~(\ref{eq:faddfullR}), leaving in the
Lehmann representation of the Faddeev propagator only those poles
$\varepsilon^{Fd}_m$, that correspond to correlated
2p1h states.
This is most easily demonstrated by applying the
the DRPA equations to both sides of Eq.~(\ref{eq:FaddTDA}).

 The vectors $\beta^{(i),m}_{\mu n_\mu \nu  n_\nu \lambda k_\lambda}$
represent the amplitudes of the three Faddeev components and sum up
to the residues of the full propagator
\begin{equation}
 b ^m_{\alpha n_\alpha \beta n_\beta \gamma  k_\gamma} ~=~ \sum_{i=1,2,3}
        \beta^{(i),m}_{\alpha n_\alpha \beta n_\beta \gamma  k_\gamma} \; .
 \label{eq:Bvsbeta}
\end{equation}
 We now define new Faddeev amplitudes $x^{(i),m}_{n_1 n_2 k}$ which are
related to the $\beta^{(i)}$'s in such a way that~\cite{Smain}
\begin{equation}
     \beta^{(i),m}_{\alpha n_1 \beta n_2 \gamma k} ~=~ 
            {{\cal X}^{n_1}_{\alpha} {\cal X}^{n_2}_{\beta} {\cal Y}^{k}_{\gamma}} 
               ~ x^{(i),m}_{n_1 n_2 k}~,
\label{eq:defx_i}
\end{equation}
where no summation is performed over the particle and hole indices $n_1$, $n_2$
and $k$.
We also introduce the notation for the spectroscopic amplitude, analogous to
Eq.~(\ref{eq:Bvsbeta})
\begin{equation}
 x^{m}_{n_1 n_2 k} ~=~  \sum_{i=1,2,3}  x^{(i),m}_{n_1 n_2 k} \; . 
\label{eq:xvsxi}
\end{equation}

 In general, $x_{n_1 n_2 k}$ and the components $x^{(i)}_{n_1 n_2 k}$
define four vectors  ${\bf x}$ and ${\bf x}^{(i)}$ all belonging to the
same linear space.
It is useful to split up the latter in two spaces $V_A$ and $V_S$
containing all the vectors that are antisymmetric and symmetric with
respect to the exchange of the two particle indices $n_1$ and $n_2$, respectively.
 Thus,
\begin{equation}
 {\bf x}\; , \; {\bf x}^{(i)}  ~ \in ~ V_A \otimes V_S \; .
\label{eq:xvctdef}
\end{equation}
We also define a vector ${\bf X}$ containing all the
three components
\begin{equation}
  {\bf X}  ~=~  \left( \begin{array}{c} {\bf x}^{(1)}  \\
                                        {\bf x}^{(2)}  \\
                                        {\bf x}^{(3)}  \end{array} \right)
  ~ \in ~ V_A^3 \otimes V_S^3 \; .
\label{eq:def-bigX}
\end{equation}
Here and in the following, we use the convention to denote vectors with
lower case boldface and operators (matrices) with plain capital letters belonging 
to the space $V_A \otimes V_S$.
Both vectors and matrices in the space $V_A^3 \otimes V_S^3$ are denoted
by capital boldface letters.
We will also use $I$ for the identity matrix in the
$V_A \otimes V_S$ space and the superscript $ex$ to indicate the vectors
obtained by exchanging the two particle indices (thus, $I^{ex}$
is the operator exchanging $n_1$ and $n_2$ in Eqs.~(\ref{eq:defx_i})
and~(\ref{eq:xvsxi})~).

\subsection{Faddeev Hamiltonian}
\label{sub:fham}

The eigenvalue equation for the Faddeev expansion can be obtained by
substituting the Lehmann representation~(\ref{eq:Fd_comp_Lehrp}) into 
Eqs.~(\ref{eq:FaddTDA}) and extracting the residues $\varepsilon^{Fd}_m$
of the poles.
 After some algebra, one obtains the following set of equations in terms of the
${\bf x}^{(i)}$ vectors
\begin{equation}
  {\bf x^{(i)}}  ~=~  \left[ H^{(i)} {H^{(i)}}^\dag  ~+~ 
                    U^{(i)} \frac{1}{\omega - D^{(i)}} {T^{(i)}}^\dag \right]
                  ~ \left(  {\bf x^{(j)}} ~+~ {\bf x^{(k)}}  \right)  
                    ~ ~ ~   i = 1, 2, 3 \; \; .
\label{eq:Xeigv1}
\end{equation}
 In Eq.~(\ref{eq:Xeigv1}), the components of the matrices $H^{(i)}$,
$U^{(i)}$ and $T^{(i)}$ are related to the spectroscopic amplitudes of the
DRPA propagators, as explained in Appendix~\ref{app:DRPA}. The
$D^{(i)}$'s are diagonal matrices containing the eigenvalues of the
corresponding DRPA.

One can now define block-diagonal matrices $\bf H$, $\bf D$, etc... which
contain on the diagonal the matrices $H^{(i)}$,
$D^{(i)}$,~etc.
These matrices act on the vectors ${\bf X}$ defined in~(\ref{eq:def-bigX}).
Using this notation one combine Eqs.~(\ref{eq:Xeigv1}) as follows
\begin{equation}
  {\bf X}  ~=~  \left[ {\bf H} {\bf H}^\dag  ~+~ 
                    {\bf U} \frac{1}{\omega - {\bf D}} {{\bf T}}^\dag \right]
                  ~  {\bf M} ~ {\bf X}   \; ,
\label{eq:Xeigv2}
\end{equation}
where we have also introduced the matrix
\begin{equation}
  {\bf M} ~=~ \left[ \begin{array}{ccc}
                     &  \; I  &  \; I \\ 
          \; I  &             &  \; I \\ 
          \; I  &  \; I  &  
                           \end{array} \right]
\label{eq:defM}
\end{equation}
that takes  into account the proper mixing between the Faddeev components.

 By introducing the vector 
\begin{equation}
  {\bf Y} ~\equiv~ \frac{1}{\omega - {\bf D}} {{\bf T}}^\dag
                  ~  {\bf M} ~ {\bf X}  
\label{eq:defY}
\end{equation}
 (which appears in Eq.~(\ref{eq:Xeigv2})~) and remembering that $\bf{D}$ is a
diagonal matrix, it is possible to manipulate Eq.~(\ref{eq:Xeigv2}) into the usual
form of an eigenvalue equation
\begin{equation}
  \omega ~ {\bf X} ~=~ {\bf F} ~ {\bf X}  
\label{eq:XeigvF}
\end{equation}
where we have introduced the Faddeev hamiltonian $\bf{F}$~\cite{Glokle99},
which 
is given by
\begin{equation}
  {\bf F} ~=~ \left[ \bf{I} - \bf{H} \bf{H}^\dag \bf{M} \right] ^{-1}
              ~ \bf{U} ~   \left\{  \bf{T}   \bf{M}   ~+~
                   \bf{D} (\bf{U}^{-1}) 
                       \left[ \bf{I} - \bf{H} \bf{H}^\dag \bf{M} \right]
                    \right\}  \; .
\label{eq:defF}
\end{equation}
 The form~(\ref{eq:XeigvF}) of the Faddeev eigenvalue equations is
useful, since it reduces the problem to the diagonalization of
a single (non-hermitian) hamiltonian.
 
 The hamiltonian $\bf{F}$ can still correspond to a large matrix requiring
a large amount of CPU time to diagonalize.
 As will be explained later in Sec.~\ref{sec:spurious}, about $2/3$ of
its solutions are trivial and without physical meaning. Thus, it is not necessary
to diagonalize the full Faddeev hamiltonian~(\ref{eq:defF}) but one can
project it onto the space of physical solutions.

\subsection{Symmetry Requirements and Normalization Conditions}

 As a consequence of the Pauli exclusion principle, the spectroscopic
amplitudes for the 2p1h motion have to be antisymmetric with respect to the
exchange of the two particle indices. This statement applies to the
full spectroscopic amplitudes~(\ref{eq:Bvsbeta}) and~(\ref{eq:xvsxi}) but not
to the single Faddeev components, which have more complicated exchange
properties.
 To exhibit the correct symmetry requirements for the Faddeev components, it is 
useful to introduce the following exchange operator, which works on the
space~(\ref{eq:def-bigX}) of the three ${\bf x}^{(i)}$ components:
\begin{equation}
  {\bf P} ~=~ \left[ \begin{array}{ccc}
                            &  \; I^{ex}  &           \\ 
          \; I^{ex}  &                    &           \\ 
                            &                    & \; I^{ex} 
                           \end{array} \right]  \; .
\label{eq:defP}
\end{equation}
 The form of matrix~(\ref{eq:defP}) takes into account that the component
$x^{(1)}$ has to change into $x^{(2)}$ when the first two legs (i.e. the two
particles) are exchanged.
 Since $\bf{P}$ is idempotent (i.e. ${\bf P}^2 = {\bf P}$), it has only
eigenvalues $+1$ and $-1$ and the respective eigenvectors are of the form
\begin{equation}
  {\bf X}_{-1}  ~=~  \left[ \begin{array}{c}   
                        {\bf x}_{a} \\
                                    - I^{ex} \; {\bf x}_{a}  \\
                        {\bf x}_{b} - I^{ex} \; {\bf x}_{b}
                                           \end{array} \right] \; \;
 \hbox{ and } \; \; 
  {\bf X}_{+1}  ~=~  \left[ \begin{array}{c}   
                        {\bf x}_{a} \\
                                    + I^{ex} \; {\bf x}_{a}  \\
                        {\bf x}_{b} + I^{ex} \; {\bf x}_{b}
                                           \end{array} \right] \; ,
\label{eq:Peigenv}
\end{equation}
 in which ${\bf x}_a$ and ${\bf x}_b$ are any two vectors.
One easily recognizes that the three Faddeev components of $\bf{X}_{-1}$
and $\bf{X}_{+1}$ give rise to antisymmetric and symmetric
spectroscopic amplitudes, respectively, when 
they are inserted in Eq.~(\ref{eq:xvsxi}).

 Using the symmetry properties of the interaction boxes~(\ref{eq:G2vsBS})
and the definition of ${\bf M}$~(\ref{eq:defM}),
one can show that $\bf{P}$ commutes with the matrix multiplying ${\bf X}$ in
Eq.~(\ref{eq:Xeigv2}) and therefore with the Faddeev
hamiltonian~(\ref{eq:defF}). Thus, ${\bf P}$ and $\bf{F}$ must have a common
set of eigenvalues.
The relevant eigenvectors in the present case correspond to those involving
$\bf{X}_{-1}$.

 The normalization condition is derived as usual by considering the
Lehmann representation~(\ref{eq:Fd_comp_Lehrp}) for the components
$R^{(i)}_{\mu \nu \lambda , \alpha \beta \gamma}(\omega)$%
~\cite{norm1,norm2}.
One can expand around a given pole $\varepsilon^{Fd}_m$ and
consider terms to order zero
and then make use of the conjugate of the eigenvalue
equation~(\ref{eq:Xeigv1}).
The result is a condition for the ${\bf X}^{(i)}$'s which only allows
proper normalization for the antisymmetric component.
These antisymmetric solutions $\bf{X}_{-1}$ 
satisfy the following condition
\begin{equation}
    {\bf x}^\dag {\bf x}  
         ~-~  \sum_{i = 1, 2, 3} {{\bf y}^{(i)}}^\dag {\bf y}^{(i)} = 2 \; ,
\label{eq:normcond}
\end{equation}
where ${\bf x}$ is the spectroscopic amplitude appearing in 
Eq.~(\ref{eq:xvsxi}) and
the factor of $2$ appears because a sum over all indices of
$x^m_{n_1 n_2 k}$~(\ref{eq:defx_i}) is implied, which includes also the
exchange terms.
 Eq.~(\ref{eq:normcond}) differs from the usual normalization of a wavefunction
for the fact that we have to subtract the additional terms
\begin{equation}
  {\bf y}^{(i)}  ~=~  \left[ V^{(i)} {H^{(i)}}^\dag  ~+~ 
                    J^{(i)} \frac{1}{\omega - D^{(i)}} {T^{(i)}}^\dag \right]
                  ~ \left(  {\bf x}^{(j)} ~+~ {\bf x}^{(k)}  \right)  \; 
                   i=1,2,3 .
\label{eq:normcond_y}
\end{equation}
These contributions 
correspond to the diagrams shown in Fig.~\ref{fig:bfflip} which have
been discarded in the present expansion.

%%%%%%%%%%%%%%%%%%%%%%%%%%%%%%%%%

\section{Treatment of Spurious solutions}
\label{sec:spurious}

 The Faddeev formalism is based on the introduction of different
components~${\bf x}^{(i)}$, which belong to the same linear space
of the total spectroscopic amplitude~${\bf x}$~(\ref{eq:xvctdef}).
These components are the solutions of the Faddeev-eigenvalue 
equation~(\ref{eq:XeigvF}), which is formulated in a larger space
in terms of the vectors ${\bf X}$ containing all three~${\bf x}^{(i)}$.
Only one third of the solutions in this larger space
have physical meaning while the others have to be discarded.
One can clarify this problem by looking at how the complete
spectroscopic amplitudes ${\bf x}$ are obtained from the components
${\bf x}^{(i)}$ through Eq.~(\ref{eq:xvsxi}). Relevant details for
treating this issue are discussed below.

 The antisymmetric solutions of the Faddeev equations
${\bf X}_{-1}$ are determined from two independent
vectors ${\bf x}_a$ and ${\bf x}_b$ as shown in Eq.~(\ref{eq:Peigenv}).
In particular, one has to specify both the symmetric and antisymmetric parts
of the first vector (${\bf x}_a$) and only the antisymmetric part
of the seocnd (${\bf x}_b$).
 These solutions therefore belong to the space
\begin{equation}
  V_F ~\equiv~ V_A ~\otimes~ V_A ~\otimes~ V_S \; .
\label{eq:defVF}
\end{equation}
The complete spectroscopic amplitudes ${\bf x}$ must also be antisymmetric
under the exchange of the two particle indices, so they belong to $V_A$.
 Thus, Eq.~(\ref{eq:xvsxi}) must be regarded as a projection from $V_F$
to the smaller space $V_A$ and therefore must have a nonvanishing kernel.
We denote this kernel by $V_{Sp}$ and refer to its vectors 
as {\em spurious states},~${\bf Y}_{Sp}$. Although these states satisfy the
Pauli requirements, they don't yield any contribution to the full 2p1h
propagator and have no physical meaning.
We also consider the space $V_{Ph}$, which is orthogonal to the kernel
$V_{Sp}$ and contains the antisymmetric states~${\bf Y}_{Ph}$ which generate
nonvanishing spectroscopic amplitudes~${\bf x}$. The vectors belonging to
$V_{Ph}$ produce contributions to the 2p1h propagator
and therefore in the following they will be referred as {\em physical states}.
In Appendix~\ref{app:spurious}, explicit basis sets for the $V_{Ph}$ and $V_{Sp}$
spaces are given. Obviously the combination of these two basis sets 
forms an orthogonal basis of $V_F$ and one has
$V_{Ph} \equiv V_A$ and $V_{Sp} \equiv V_A \otimes V_S$.

It must be stressed that in general
the solutions of the Faddeev eigenvalue equation~(\ref{eq:XeigvF}) do not
automatically separate into the physical and spurious states just defined. 
Nevertheless, it is shown in Appendix~\ref{app:spurious} that
the states of $V_{Sp}$ are proper eigenstates of the Faddeev
hamiltonian for the expansion presented in this paper.
This feature always occurs for the three-body problem but is not guaranteed
when working with quasiparticle and quasihole excitations unless a proper
set of diagrams is considered.
 When this condition is satisfied, there exist a set of spurious solutions of
the Faddeev equation ~(\ref{eq:XeigvF}) which spans the space $V_{Sp}$
completely.
 The projection of the Faddeev hamiltonian~(\ref{eq:defF}) onto the
physical and spurious subspaces $V_{Ph}$ and $V_{Sp}$ then takes the form
\begin{equation}
   {\bf F} ~=~  \left[ \begin{array}{cc} 
                  {\mbox{$\langle {Ph} \vert $}} {\bf F} {\mbox{$\vert {Ph} \rangle$}}  &      0      \\
                  {\mbox{$\langle {Sp} \vert $}} {\bf F} {\mbox{$\vert {Ph} \rangle$}}  &
                         {\mbox{$\langle {Sp} \vert $}} {\bf F} {\mbox{$\vert {Sp} \rangle$}}
                 \end{array} \right] 
     \; .
\label{eq:proj_F}
\end{equation}

It should be noted that the physical states ${\bf Y}_{Ph}$, belonging
to $V_{Ph}$, differ from the spurious ones ${\bf Y}_{Sp}$~($\in V_{Sp}$) not
only because they give rise to physically meaningful spectroscopic
amplitudes but also because they are not solutions of the Faddeev
equations~(\ref{eq:XeigvF}).
 In general, a physically meaningful eigenvector of~(\ref{eq:proj_F}),
${\bf X}_{physical}$, is a mixture of states belonging to both
$V_{Ph}$ and $V_{Sp}$, due to the mixing term
${\mbox{$\langle {Sp}\vert $}} {\bf F} {\mbox{$\vert {Ph} \rangle$}}$.
Thus,
\begin{equation}
 {\bf X}_{physical} ~=~ c_1 \; {\bf Y}_{Ph} + c_2 \; {\bf Y}_{Sp} ~, 
\label{eq:phsol}
\end{equation}
with $c_1$ and $c_2$ some constants.
In other words, a spurious component ${\bf Y}_{Sp}$ is also generated which
will be automatically projected out when computing the spectroscopic
amplitude~${\bf x}$~(\ref{eq:xvctdef}).

 It is important to recognize that such spurious contributions are indeed
needed since they account for the differences of the three Faddeev
components~(\ref{eq:defx_i}).
 The relation between the usual Faddeev components for a given physical or
spurious state can be inferred from the basis sets~(\ref{eq:xphandxsp}).
 There it is shown that all the Faddeev components ${\bf x}^{(i)}$
of a state ${\bf Y}_{Ph}$ or ${\bf Y}_{Sp}$ are equal up, to a sign.  
  As a consequence, if a general solution is a pure physical
state ${\bf Y}_{Ph}$, all its Faddeev components cannot differ
from each other in a significant way.
 Having a mixing between physical and spurious states allows the possibility
of obtaining two independent Faddeev components.
This result corresponds to the physical ingredients which involve identical
ph phonons for the components ${\bf x}^{(1)}$ and ${\bf x}^{(2)}$ but
a pp phonon for ${\bf x}^{(3)}$.

When all the Faddeev components are summed
to generate the full ${\bf x}$ in~(\ref{eq:xvsxi}), the contribution
of the spurious states cancels out.
 Thus, for any nonspurious solution of the Faddeev equations, only the
contribution from physical states ${\bf Y}_{Ph}$ is needed
to determine the 2p1h propagator.
 By looking at Eq.~(\ref{eq:proj_F}), it is easy to see that these
contributions can
be directly obtained by diagonalizing the upper-left block
\begin{equation}
   \omega_{m} \; {\bf Y}^{m}_{Ph} ~=~
           {\mbox{$\langle {Ph}\vert $}} {\bf F} {\mbox{$\vert {Ph}\rangle$}}
           ~ {\bf Y}^{m}_{Ph} \; ,
\label{eq:xph_diag}
\end{equation}
where $m$ is used to label the solutions.
The solutions of Eq.~(\ref{eq:xph_diag}) are sufficient
to determine the 2p1h propagator.
For some applications one may need the individual
components ${\bf x}^{(i)}$.
In that case, the contribution from
spurious states ${\bf Y}_{Sp}$  can be
determined by solving the remaining part of the Faddeev equations
\begin{equation}
   \omega_{m} \; {\bf Y}^{m}_{Sp} = 
        {\mbox{$\langle {Sp}\vert $}} {\bf F} {\mbox{$\vert {Ph}\rangle$}}
                                          ~ {\bf Y}^{m}_{Ph} 
      + {\mbox{$\langle {Sp}\vert $}} {\bf F} {\mbox{$\vert {Sp}\rangle$}}
                                          ~ {\bf Y}^{m}_{Sp} \; .
\label{eq:xsp_vs_xph}
\end{equation}

We note that if the upper-right block of
Eq.~(\ref{eq:proj_F}) is not zero, a mixing between the ${\bf Y}_{Ph}$ and
${\bf Y}_{Sp}$ states occurs for all the eigenstates of the Faddeev
hamiltonian.
In this situation, the spurious eigenvalues will differ from the unperturbed
energies and all of the solutions of the Faddeev equations will contain
a component ${\bf Y}_{Ph}$. The Faddeev formalism would therefore become useless,
since it would no longer be possible to discern between 
``good'' and ``bad'' solutions.
 In Appendix~\ref{app:spurious} we show how the correct behavior of spurious
solutions is related to the presence of backward-going contributions
of the DRPA $\Gamma$-matrices~(see Fig.~\ref{fig:Gammai_vs_DRPAS}).
 In case these diagrams are neglected, the
spurious states ${\bf Y}_{Sp}$ no longer diagonalize the Faddeev hamiltonian.
 Such diagrams may give a small contribution to the description of low-lying
states but they are essential to make the whole formalism presented
here meaningful.
 As a general rule, when deriving expansions based on the Faddeev equations, it
should be kept in mind that not all possible sets of diagrams can be
effectively summed to all orders. Instead, one must first check
the consistency of the set of diagrams
with respect to the behavior of spurious
solutions.

%%%%%%%%%%%%%%%%%%%%%%%%%%%%%%%%%%%%%%%%
\section{Summary and conclusions}
\label{sec:conclusions}

The present theoretical description of the distribution of spectroscopic strength
at low energies lacks important ingredients for a successful
comparison with experimental data.
One of these ingredients is a proper description of the coupling
of sp motion to low-lying collective modes that are present in the system.
Recent calculations for ${}^{16}{\rm O}$~\cite{GeurtsO16}, for example,
only include a TDA description of these collective modes.
 A new method is proposed here to study the influence of pp and ph RPA
correlations on the sp propagator for a system with a finite
number of fermions.
 This method is formulated in the context of SCGF theory 
by evaluating the nucleon self-energy in terms of
the 2p1h and 2h1p propagators.
 The description of the 2p1h (or 2h1p) excitations has been studied by using the
Faddeev formalism, which is usually applied to solve the three-body problem.
The Faddeev formalism is necessary since we consider the collective pp and
ph RPA phonons as the basic building blocks to describe the 2p1h motion.

 The computational scheme presented here employs only two-time
propagators, thus leading to a tractable set of equations. At the same time the
contributions of pp and ph RPA phonons have been consistently summed to all
orders thereby
including the physical effects that appear to be relevant for the study
of the ${}^{16}{\rm O}$ nucleus. Unlike previous calculations in which ph phonons
have been included, the present
formalism takes the Pauli-exchange correlations properly into account up
to the 2p1h level. 

 In deriving the set of Faddeev equations a formulation has been chosen
which involves only a single diagonalization for the 2p1h fragments.
 The appearance of spurious solutions has also been discussed in some
detail, showing that the inclusion of the contribution of
certain diagrams is necessary to separate such spurious solutions from
the physically meaningful ones.
 When this separation occurs, it is straightforward to project out the
physical eigenstates from the Faddeev equations, thereby eliminating the
spurious ones.

The Faddeev formalism has been used to include specific correlations 
corresponding to pp and ph phonons in a natural way.
Extensions to the inclusion of more complicated excitations like the
extended DRPA~\cite{DRPApaper} can be obtained in a convenient way
by starting from the formalism presented in Sec. \ref{sec:Rand4timesFd}.

The formalism presented here appears practical for describing the
spectroscopic strength in ${}^{16}{\rm O}$ in a similar space as was employed
in~\cite{GeurtsO16}.
This implementation is currently in progress and will be reported elsewhere.

%%%%%%%%%%%%%%%%%%%%%%%%%%%%%%%%%%%%%%%%%%%%
\acknowledgments
This work is supported by the
U.S. National Science Foundation under Grant No.
PHY-9900713.
%%%%%%%%%%%%%%%%%%%%%%%%%%%%%%%%%%%%%%%

%%%%%%%%%%%%%%%%%%%%%%%%%%%%%%%%%%%%%%%%%%%
%%
%%  Appendices
%%
\appendix

%%%%%%%%%%%%%%%%%%%%%%%%%%%%%%%%%%%%%%%%

\section{Interaction boxes.}
\label{app:Gboxes}

The $\tilde{\Gamma}^{(pp)}$ matrix~(\ref{eq:ppBS}) obtained by solving the
DRPA equation has the following Lehmann representation
\begin{eqnarray}
  \tilde{\Gamma}^{(pp)}_{\mu \nu , \alpha \beta}(\omega) ~&=&~
         V_{\mu \nu , \alpha \beta}  ~+~
  \sum_{n+} \frac{\left( \Delta^{n+}_{\mu \nu} \right)^* \Delta^{n+}_{\alpha \beta}  }
              { \omega - \varepsilon^{\Gamma+}_{n+} + i \eta } ~-~
  \sum_{k-} \frac{\Delta^{k-}_{\mu \nu} \left( \Delta^{k-}_{\alpha \beta} \right)^*  }
              { \omega - \varepsilon^{\Gamma-}_{k-} - i \eta }  \; ,
\nonumber  \\
  &\equiv&  ~   V_{\mu \nu , \alpha \beta}
         ~+~ \Delta \tilde{\Gamma}^{>}_{\mu \nu , \alpha \beta}(\omega) 
         ~+~ \Delta \tilde{\Gamma}^{<}_{\mu \nu , \alpha \beta}(\omega) 
\label{eq:Gpp_leh}
\end{eqnarray}
in which $n+$ ($k-$) label the forward-going (backward-going) contributions.
 In obtaining the $\Gamma^{(3)}$ vertex given by
Eq.~(\ref{eq:G2vsBS_pp}) we want to keep both the forward-
and backward-going terms of~(\ref{eq:Gpp_leh}).
This implies that all three diagrams of Fig.~\ref{fig:Gammai_vs_DRPAS}
are included.
 The main problem encountered when working with dressed propagators is that 
the contribution of these three diagrams do not factorize in an expression
of the form~${G^0}^> \Gamma {G^0}^>$ when ${G^0}^>$ is represented by
a propagator of the form
\begin{equation}
  {G^0}^>_{\mu \nu \lambda , \alpha \beta \gamma}(\omega) ~=~
     \sum_{n_1, n_2, k}
     \frac{ \left( 
         {\cal X}^{n_1}_{\mu} {\cal X}^{n_2}_{\nu} {\cal Y}^{k}_{\lambda}
            \right)^* \; 
        {\cal X}^{n_1}_{\alpha} {\cal X}^{n_2}_{\beta} {\cal Y}^{k}_{\gamma} }
    { \omega - ( \varepsilon^+_{n_1} + \varepsilon^+_{n_2} -
                                     \varepsilon^-_{k} ) + i \eta } \; ,
\label{eq:G0fw_app}
\end{equation}
 This factorization cannot be made because of 
the implicit sums over the particle and hole excitation
indices~$(n,n',k)$ in Eq.~(\ref{eq:G0fw_app}).
For example, the hole label $k$ cannot change in 
Fig.~\ref{fig:Gammai_vs_DRPAS}~\cite{jyuan,DRPApaper,Geurtsth}.

 This difficulty can be overcome with a slight reformulation of the
problem.  We no longer regard $\Gamma^{(3)}$ and ${G^0}^>$ only as
functions of the model space indices $(\alpha,\beta,\gamma)$, but instead
assign an additional dependence on the particle and hole indices.
Thus promoting the $(n,n',k)$ quantum
numbers to external indices, the Lehmann representation of
${G^0}^>$~(\ref{eq:G0fw}) will contain at most one pole for every matrix
element.
As a consequence, all the components~(\ref{eq:faddcmpall}) 
appearing in the Faddeev equations
have to be reformulated in the same way. The original propagators
can then be retrieved at the end by summing the solutions over all the
particle and hole fragments.
With this procedure it becomes possible to write
the sum of the three diagrams in Fig.~\ref{fig:Gammai_vs_DRPAS} 
in terms a matrix product of two ${G^0}^>$~(\ref{eq:G0fw}) propagators and
the following vertex
\begin{eqnarray}
\lefteqn{
  \Gamma^{(3)}_{\mu n_\mu \nu n_\nu \lambda k_\lambda ,
           \alpha n_\alpha \beta n_\beta \gamma k_\gamma}(\omega) ~=~
        {\textstyle {1\over 2}} \frac{\delta_{k_\lambda , k_\gamma}}
                 { \sum_\sigma \left| {\cal Y}^{k_\lambda}_\sigma \right|^2 } }
    \hspace{.5in} & &
\nonumber \\
 ~ ~ ~ ~ ~ ~ ~ ~ ~ ~ ~ ~
 & & \times ~ \left\{      V_{\mu \nu , \alpha \beta}  ~+~
       \sum_{n+} 
         \frac{\left( \Delta^{n+}_{\mu \nu} \right)^* \Delta^{n+}_{\alpha \beta}  }
              { \omega - ( \varepsilon^{\Gamma+}_{n+} - \varepsilon^-_{k_\lambda} ) + i \eta } \right.
\nonumber \\
 & & ~+~ \left.   \sum_{k-} 
         \frac{ [ \omega - \varepsilon^+_{n_\mu} - \varepsilon^+_{n_\nu}
                          - \varepsilon^+_{n_\alpha} - \varepsilon^+_{n_\beta}
                           + \varepsilon^-_{k_\lambda} + \varepsilon^{\Gamma-}_{k-} ]
        \;  \Delta^{k-}_{\mu \nu} \left( \Delta^{k-}_{\alpha \beta} \right)^*  }
              { (\varepsilon^{\Gamma-}_{k-} - \varepsilon^+_{n_\mu} - \varepsilon^+_{n_\nu} )
                (\varepsilon^{\Gamma-}_{k-} - \varepsilon^+_{n_\alpha} - \varepsilon^+_{n_\beta} ) }
 \right\} \; ,
\label{eq:Gamma3}
\end{eqnarray}
which corresponds to the expression for the pp interaction
box~(\ref{eq:G2vsBS_pp}).
 With this prescription, we are able to write an expansion
that sums diagrams like those of Fig~\ref{fig:Gammai_vs_DRPAS}. This is
achieved at the cost of an increased size of the matrices to be dealt with.
 After further manipulation, it is possible
to avoid this complication by dropping the dependence on the model
space indices $(\alpha,~\beta,~\gamma)$, as explained in Sec.~\ref{sub:xamp}.
The advantage of the present procedure lies in the possibility to
diagonalize the Faddeev amplitudes in one step instead of solving the equations
with energy-dependent vertex functions as discussed in Sec.~\ref{sub:fham}.
 The expressions for the ph interaction boxes $\Gamma^{(1)}$ and
$\Gamma^{(2)}$ are derived in a completely analogous way.

%%%%%%%%%%%%%%%%%%%%%%%%%%%%%%%%%%%%%%%%
\section{Dressed RPA equations.}
\label{app:DRPA}

To clarify the notations used in the paper, we give here a brief
overview of the DRPA equation for the pp interaction matrix.
 We also give the explicit expressions for the normalization and closure
relations used in the development of the formalism.

 The pp-DRPA equation is derived from Eq.~(\ref{eq:ppBS}) by choosing
$K^{(pp)}_{\alpha \beta , \gamma \delta} = V_{\alpha \beta , \gamma \delta}$ 
and is shown in Fig.~\ref{fig:Gm_n_pro-pp}.
 Using the Lehmann representation~(\ref{eq:Gpp_leh}) and extracting the poles
$\varepsilon^{\Gamma+}_{n+}$~($\varepsilon^{\Gamma-}_{k-}$) from the DRPA
equation, we get the usual eigenvalue problem
\begin{equation}
 \Delta^{n+(k-)}_{\alpha \beta} ~=~ \left.  \Delta^{n+(k-)}_{\gamma \delta} ~
           g^{(0)}_{\gamma \delta , \epsilon \rho}(\omega)~
           {\textstyle {1\over 2}} ~ V_{\epsilon \rho ,\alpha \beta}
           \right| _{ \omega = \varepsilon^{\Gamma+}_{n+}~
                            (\varepsilon^{\Gamma-}_{k-}) }   \; ,
\label{eq:ppLd_eig}
\end{equation}
where $n+$ ($k-$) refer to the forward-going (backward-going) solutions.
 It is useful to introduce the following notation, in analogy
to the convention introduced in Eq.~(\ref{eq:defx_i}) for the Faddeev
components,
\begin{mathletters}
\label{def:ld_comp}
\begin{equation}
  U^{n+}_{n_1,n_2} ~=~   \frac{ {\cal X}^{n_1}_{\alpha} {\cal X}^{n_2}_{\beta} 
                       ~ \left(   \Delta^{n+}_{\alpha \beta}  \right)^\ast }
       { \sqrt{2} ~ ( \varepsilon^{\Gamma+}_{n+} - \varepsilon^+_{n_1} - \varepsilon^+_{n_2} )  }  \; ,
\label{def:ld_compU}
\end{equation}
\begin{equation}
\label{def:ld_compH}
  H^{k-}_{n_1,n_2} ~=~  \frac{ {\cal X}^{n_1}_{\alpha} {\cal X}^{n_2}_{\beta} 
                         ~  \Delta^{k-}_{\alpha \beta} }
       { \sqrt{2} ~ ( \varepsilon^{\Gamma-}_{k-} - \varepsilon^+_{n_1} - \varepsilon^+_{n_2} )  }  \; ,
\end{equation}
\begin{equation}
\label{def:ld_compJ}
 J^{n+}_{k_1,k_2} ~=~  \frac{\left(  {\cal Y}^{k_1}_{\alpha} {\cal Y}^{k_2}_{\beta}
                          ~   \Delta^{n+}_{\alpha \beta} \right)^\ast }
       { \sqrt{2} ~ ( \varepsilon^{\Gamma+}_{n+} - \varepsilon^-_{k_1} - \varepsilon^-_{k_2} )  }  \; ,
\end{equation}
\begin{equation}
\label{def:ld_compV}
  V^{k-}_{k_1,k_2} ~=~  \frac{\left(  {\cal Y}^{k_1}_{\alpha} {\cal Y}^{k_2}_{\beta}
                       \right)^\ast   ~ \Delta^{k-}_{\alpha \beta} }
       { \sqrt{2} ~ ( \varepsilon^{\Gamma-}_{k-} - \varepsilon^-_{k_1} - \varepsilon^-_{k_2} )  }  \; .
\end{equation}
\end{mathletters}
These represents the generalization to dressed propagators of the usual RPA
components (the $\sqrt{2}$ has been inserted only in the pp case for
convenience). 
In Eq.~(\ref{def:ld_comp}) the quantities~${\cal{X}}$~(${\cal{Y}}$) and%
~$\varepsilon^+_n$~($\varepsilon^-_k$) represent the spectroscopic amplitudes and the poles of
the forward-going (backward-going) part of the one-body propagator, while
$\varepsilon^{\Gamma+(-)}_{n+(k-)}$ are the eigenvalues of the DRPA
equation~(\ref{eq:ppLd_eig}).

 The normalization condition for the DRPA solutions, given in terms of the
components~(\ref{def:ld_comp}), is the generalization of the normalization
for the usual RPA~\cite{schuckbook} and can be put in matrix notation as
\begin{equation}
\label{eq:ld_norm}
  \left[ \begin{array}{cc}  U^\dag  &  J^\dag \\ 
                        H^\dag & V^\dag  \end{array} \right]
  \left[ \begin{array}{cc}  \; I  &   \\ 
                                    &  - I  \end{array} \right]
  \left[ \begin{array}{cc}  U  &  H \\ 
                            J & V  \end{array} \right]
 ~=~ \left[ \begin{array}{cc}  \; I  &   \\ 
                                 & - I  \end{array} \right]  \; ,
\end{equation}
while the closure relations are given by
\begin{equation}
\label{eq:ld_closure}
  \left[ \begin{array}{cc}  U  &  H \\ 
                            J & V  \end{array} \right]
  \left[ \begin{array}{cc}  \; I  &   \\ 
                                    &  - I  \end{array} \right]
  \left[ \begin{array}{cc}  U^\dag  &  J^\dag \\ 
                        H^\dag & V^\dag  \end{array} \right]
 ~=~ \frac{1}{2} \left[ \begin{array}{cc}  \; I - I^{ex}  &   \\ 
                                 & I^{ex} - I  \end{array} \right]  \; ,
\end{equation}
 where $U$, $H$, $J$ and $V$ are the matrices containing the
elements of Eq.~(\ref{def:ld_comp}).
 In dealing with the formalism for the Faddeev equations, it is also useful
to introduce the following two matrices:
\begin{mathletters}
\label{def:ld_compTW}
\begin{equation}
  T^{n+}_{n_1,n_2} ~=~   \frac{1}{\sqrt{2}}
       {\cal X}^{n_1}_{\alpha} {\cal X}^{n_2}_{\beta} 
                       ~ \left(   \Delta^{n+}_{\alpha \beta}  \right)^\ast  \; ,
\label{def:ld_compT}
\end{equation}
\begin{equation}
  W^{k-}_{k_1,k_2} ~=~  \frac{1}{\sqrt{2}}
        \left(  {\cal Y}^{k_1}_{\alpha} {\cal Y}^{k_2}_{\beta}
                       \right)^\ast   ~ \Delta^{k-}_{\alpha \beta}  \; .
\label{def:ld_compW}
\end{equation}
\end{mathletters}
which are trivially related to the components~(\ref{def:ld_compU})
and~(\ref{def:ld_compV}).

 The matrix elements given in Eqs.~(\ref{def:ld_comp})
and~(\ref{def:ld_compTW}) correspond to the matrices $H^{(3)}$,
$U^{(3)}$, and $T^{(3)}$ introduced after Eq.(\ref{eq:Xeigv1}) for the
2p1h Faddeev expansion ($J$, $V$, and $W$ being the
corresponding ones for the 2h1p expansion).

%%%%%%%%%%%%%%%%%%%%%%%%%%%%%%%%%%%%%%%%

\section{Properties of Spurious States}
\label{app:spurious}

 The set of solutions ${\bf X}_{-1}$~(\ref{eq:Peigenv}) which satisfy the
Pauli requirements can be divided in two subsets of physical $V_{Ph}$ and
spurious states $V_{Sp}$. Orthogonal basis sets for these two spaces are
given by
\begin{equation}
  {\bf Y}_{Ph} ~=~ \left( \begin{array}{c}
                        {\bf u} - {\bf u}^{ex} \\
                        {\bf u} - {\bf u}^{ex} \\
                        {\bf u} - {\bf u}^{ex}
                       \end{array} \right)   \in V_{Ph}
 \; \;   \hbox{ and } \; \;
  {\bf Y}_{Sp} ~=~  \left( \begin{array}{c}
                      - {\bf u}                \\
                                + {\bf u}^{ex} \\
                        {\bf u} - {\bf u}^{ex}
                       \end{array} \right)   \in V_{Sp} \; .
\label{eq:xphandxsp}
\end{equation}
 where the ${\bf u}$ represent unit vectors which belong to the
space~(\ref{eq:xvctdef}). Their
components are given by
\begin{equation}
 u_{n_1 n_2 k} ~=~  \delta_{n_1 , n'} \delta_{n_2 , n''} \delta_{k , k'}
\label{eq:def_u}
\end{equation}
with $n'$, $n''$ and $k'$ fixed fragmentation indices which label all the 
possible ${\bf u}$.   The vectors ${\bf u}^{ex} = I^{ex} ~ {\bf u}$ are
given by the exchange of the two particle indices $n_1$ and $n_2$.
The physical states ${\bf Y}_{Ph}$ are characterized by the fact
that they do not produce vanishing spectroscopic amplitudes while the spurious
states ${\bf Y}_{Sp}$ do. Thus $V_{Sp}$ represents the kernel
of Eq.~(\ref{eq:xvsxi}).

 It is clear from Eq.~(\ref{eq:xphandxsp}) that the ${\bf Y}_{Ph}$ states
span a space equivalent to the space of antisymmetric vectors
{\bf x}~(\ref{eq:xvsxi}), thus $V_{Ph} \equiv V_A$.
 Analogously, the ${\bf Y}_{Sp}$ states depend on both the symmetric and the
antisymmetric parts of the ${\bf u}$ vectors, which implies
$V_{Sp} \equiv V_A \otimes V_S$.
 Therefore, the vectors~(\ref{eq:xphandxsp}) form a basis for the full
antisymmetric Faddeev space $V_F$~(\ref{eq:defVF}).

 In general, the physical and spurious states~(\ref{eq:xphandxsp}) defined
here are not solutions of the Faddeev equations~(\ref{eq:XeigvF}), they
simply define a basis over which these solutions can be expanded.
 Nevertheless, for both the normal three-body Faddeev equations and
the expansion proposed in this paper, it can be seen that the spurious states
${\bf Y}_{Sp}$ (and only those) diagonalize the Faddeev hamiltonian. The 
eigenvalues correspond to the poles of the three freely propagating lines%
~$\omega = \varepsilon^{+}_{n'} + \varepsilon^{+}_{n''} -\varepsilon^{-}_{k}$%
~(\ref{eq:G0fw}).
 This feature serves as a sum rule on the solutions of the Faddeev equations
and (unlike the case of 3-body systems) is not always satisfied when 
applying the formalism to particle and hole excitations.
 Instead this property depends on the diagrams included in the expansion and 
a proper set of  diagrams needs to be employed in order to apply
the Faddeev formalism.
 For the particular Faddeev expansion described here, this constraint
is achieved by including the backward-going terms of DRPA phonons
in the $\Gamma^{(i)}$ matrices and by using the closure
relations~(\ref{eq:ld_closure}), which turn out to play an important role.
In the following, the proof that the ${\bf Y}_{Sp}$ states
of Eq.~(\ref{eq:xphandxsp}) actually represent a set of spurious solutions
of the present Faddeev formalism is outlined. 
 This also clarifies the relationship between
the correct behavior of the spurious solutions and the backward-going DRPA
diagrams.

 Consider a spurious state $\tilde{\bf Y}_{Sp}$ of the
form~(\ref{eq:xphandxsp}), with $\tilde{\bf u}$ given by Eq.~(\ref{eq:def_u})
and eigenvalue $\tilde{\omega}$.
 We now observe that the matrices $U$~(\ref{def:ld_compU}) and
$T$~(\ref{def:ld_compT}) differ from each other only by an energy
denominator. In particular, we have
\begin{equation}
 - \frac{1}{\tilde{\omega} - {\bf D}} \; {\bf T}
  \rightarrow {\bf U} \; ,
\label{eq:UvsT}
\end{equation}
were ${\bf U}$, ${\bf T}$ and ${\bf D}$ are defined in Sec.~\ref{sec:Faddeev}.
 If the eigenvalue is given by
$\tilde{\omega}=\varepsilon^{+}_{n'}+\varepsilon^{+}_{n''}-\varepsilon^{-}_{k}$,
the equivalence of the left- and right-hand side holds {\em only} for the matrix
elements having the same indices~$(n',n'',k')$.
Indeed, only in that case the denominator
$\tilde{\omega} - {\bf D}$ will be equal to the one in
Eq.~(\ref{def:ld_compU}).
 On the other hand, we see from Eq.~(\ref{eq:def_u}) that the
components of $\tilde{\bf u}$ are nonzero only for the same indices.
 This allows the substitution of the $\rightarrow$ in Eq.~(\ref{eq:UvsT})
by an equal sign when
acting on the vector~$\tilde{\bf Y}_{Sp}$.
 Substituting Eq.~(\ref{eq:UvsT}) into Eq.~(\ref{eq:Xeigv2}) and using the 
closure relations of the DRPA, we obtain the equation
\begin{equation}
 \tilde{\bf Y} _{Sp} ~=~ - \; {\bf M}  \; \tilde{\bf Y} _{Sp}  \; \; ,
\label{eq:X-MX}
\end{equation}
which is valid only for the specific state $\tilde{\bf Y}_{Sp}$, 
labeled by the indices~$(n',n'',k')$.  The last equation is satisfied
for a spurious states of the form~(\ref{eq:xphandxsp}) but not for 
the corresponding physical state~$\tilde{\bf Y}_{Ph}$.
 Thus, we have obtained a set of spurious solutions of the Faddeev
equations which form an orthogonal  basis of~$V_{Sp}$.

 In this proof, we note that the closure relation~(\ref{eq:ld_closure}) can be
applied to derive Eq.(\ref{eq:X-MX}) because of the presence of the
backward-going term ${\bf H}{\bf H}^\dag$ in Eq.~(\ref{eq:Xeigv2}), which comes
directly from the last diagram of Fig.~\ref{fig:Gammai_vs_DRPAS}.

%%%%%%%%%%%%%%%%%%%%%%%%%%%%%%%%%%%%%%%%%%%%

%%
%% FIGURES
%%
%%%%%%%%%%%%%%
%%%%%%%%%%%%%%

\begin{figure}
 \begin{center}
 \end{center}
\caption[]{Example of an approximation for the self-energy. Although this
approximation
contains both ph and pp correlations it would generate incorrect
results, due to
the need of subtracting the 2nd order term to avoid double counting.
\label{fig:pp_ph_sf} }
\end{figure}

\begin{figure}
 \begin{center}
 \end{center}
\caption[]{Bethe-Salpeter equation~(\ref{eq:Rbt}) for the 2p1h 
propagator with vertices given by~(\ref{eq:Rbtvertex}). 
 The irreducible interaction vertices for the pp and two ph channels
 are denoted by $K^{pp}$ and $K^{ph}$, respectively.
The irreducible vertex involving all three lines simultaneously is denoted by
$K^{pph}$.
\label{fig:BS2p1h} }
\end{figure}

\begin{figure}
 \begin{center}
 \end{center}
\caption[]{Diagrams that are included in the definition of the vertex
for the pp channel.
 Here $\Delta \Gamma^>$ and $\Delta \Gamma^<$ are the forward- and
backward-going part of the energy dependent contribution to the pp DRPA
vertex~(\ref{eq:Gpp_leh}).
  The contribution of these three diagrams can be 
factorized in an expression of the form ${G^0}^> \; \Gamma^{(3)} \; {G^0}^>$ 
only after having redefined the propagators ${G^0}^>$ and $\Gamma^{(3)}$
to depend also on the particle and hole
fragmentation indices ($n$,$n'$,$k$).
 The last diagram has a smaller effect on the physical solutions of the
problem, although it is essential for the elimination of spurious solutions.
\label{fig:Gammai_vs_DRPAS} }
\end{figure}

\begin{figure}
 \begin{center}
 \end{center}
\caption[]{A combination of two diagrams of the type shown here
can be used to connect the 2p1h and 2h1p propagators.
Diagrams like these are not included in the present 
approximation scheme.
Nevertheless, their contribution appears in the normalization of
spectroscopic amplitudes.
Explicit time-ordering is implied in this diagram.
\label{fig:bfflip} }
\end{figure}

\begin{figure}
 \begin{center}
 \end{center}
\caption[]{DRPA equation for the $\tilde{\Gamma}^{(pp)}$ matrix.
\label{fig:Gm_n_pro-pp} }
\end{figure}

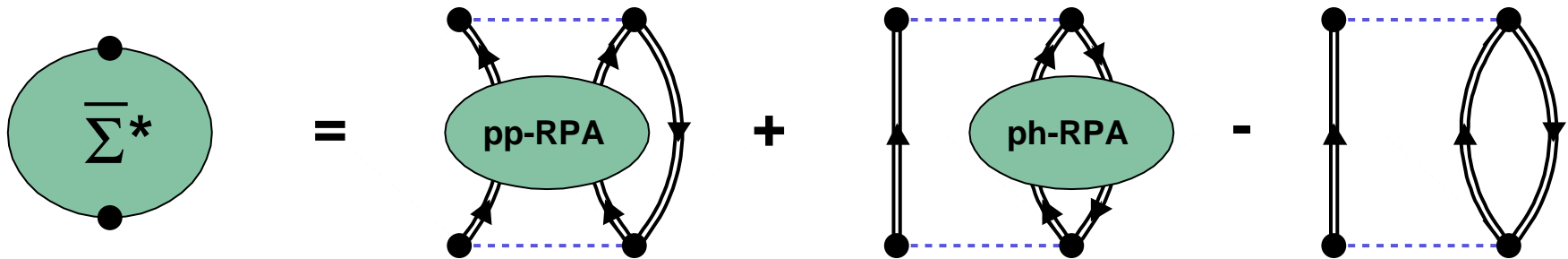
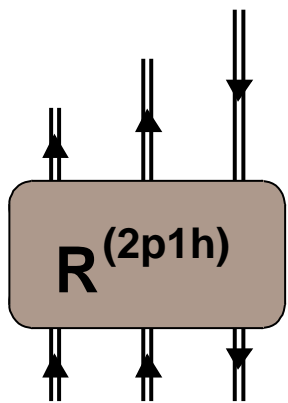
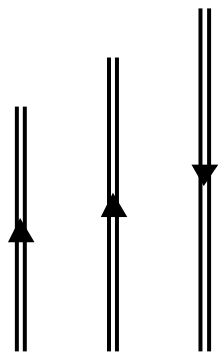
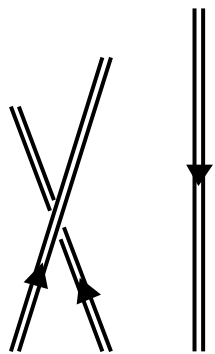
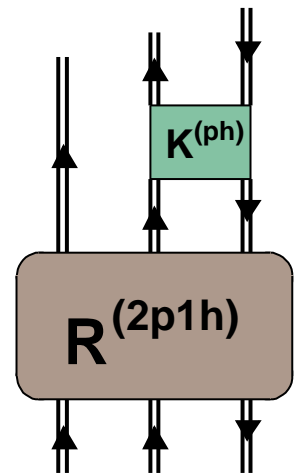
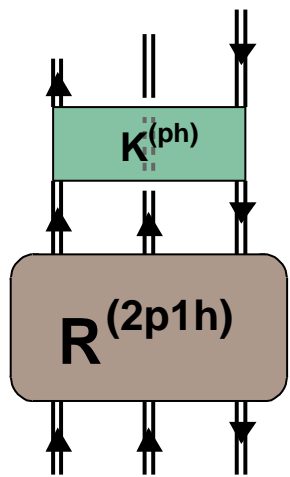
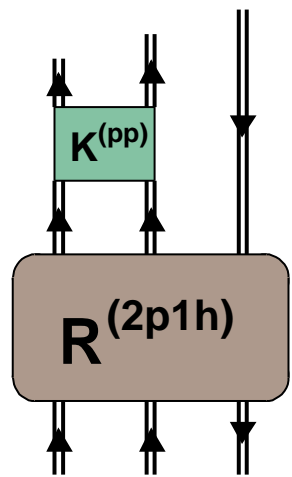
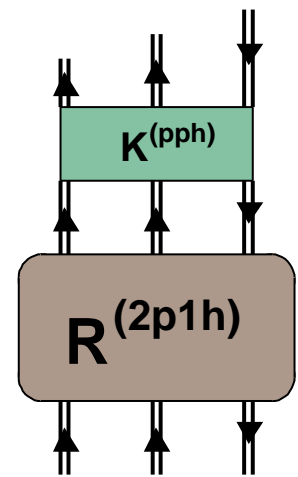
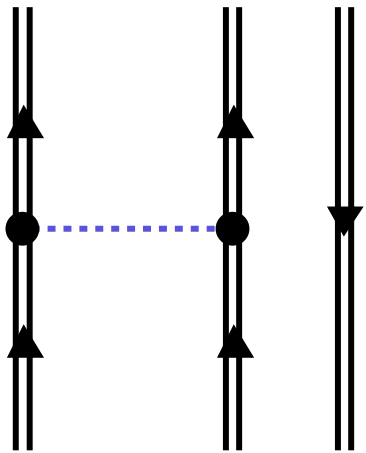
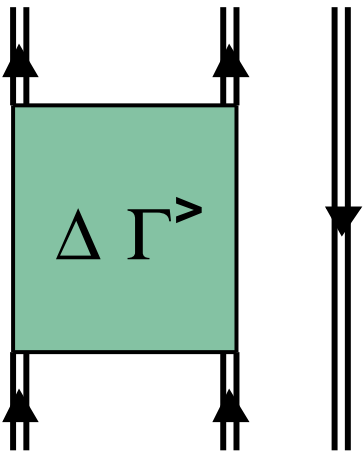
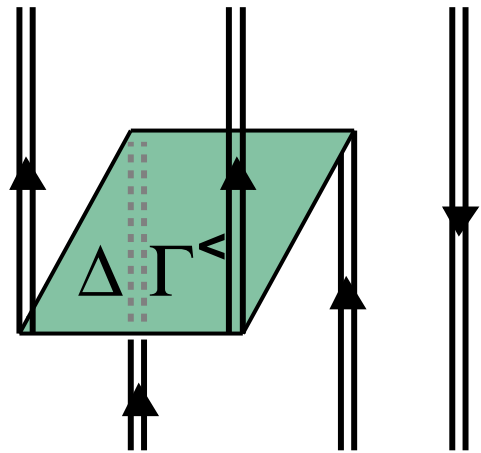
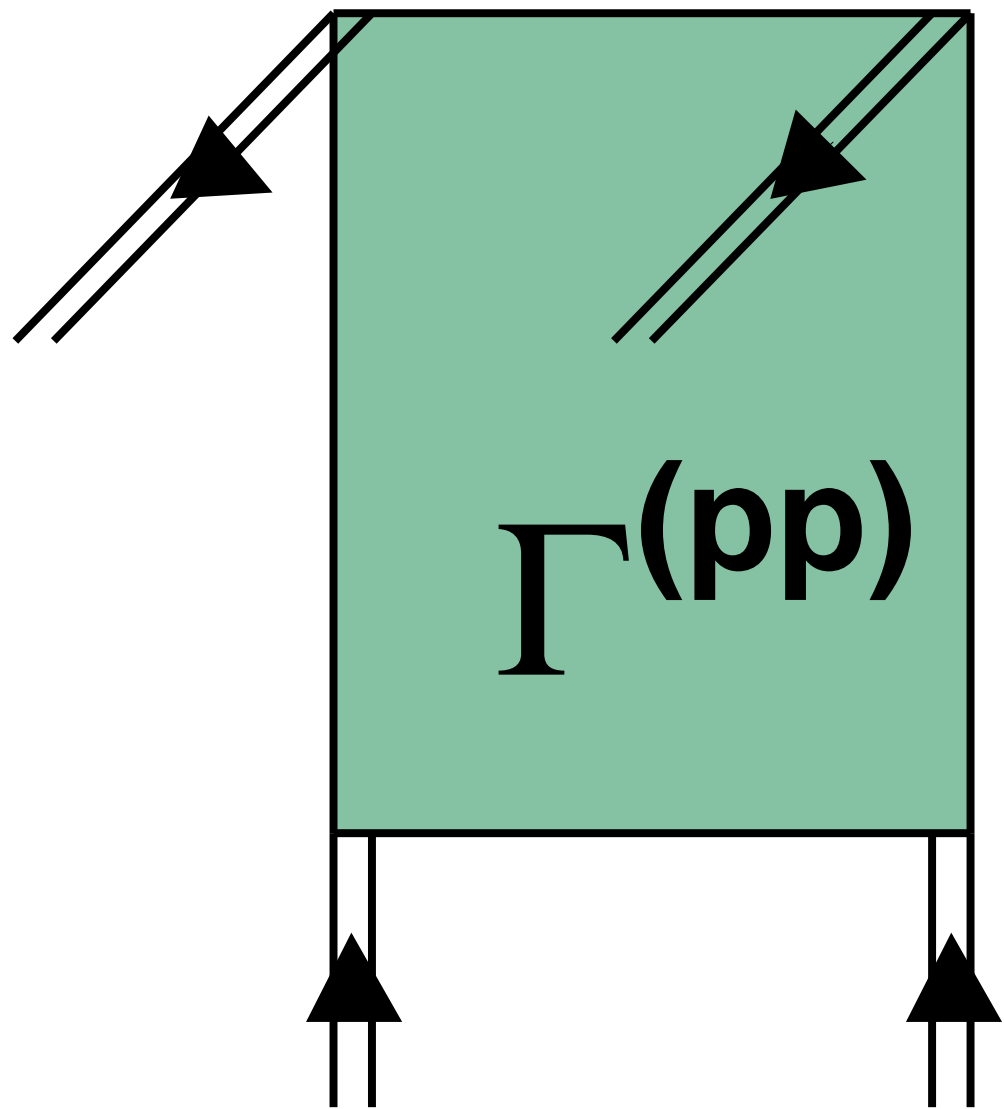
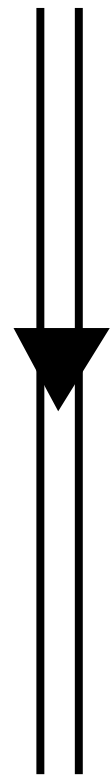
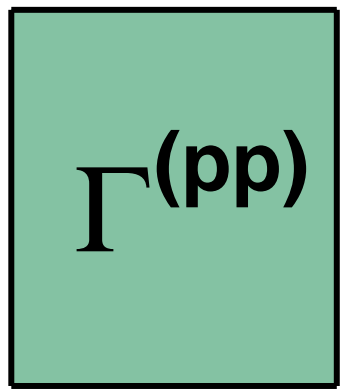
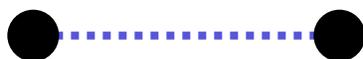
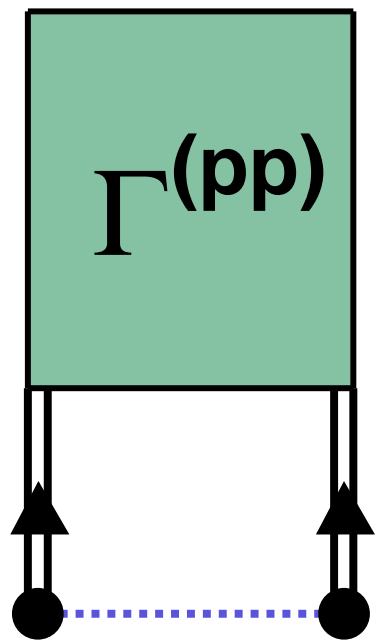
\end{document}